\documentclass[preprint]{iucr}              % DO NOT DELETE THIS LINE
\usepackage{array}
\usepackage{url}
\usepackage{hyperref}
\usepackage{color}
\usepackage{amssymb}
\usepackage{xcolor}
     \journalcode{J}              % Indicate the journal to which submitted
\begin{document}                  % DO NOT DELETE THIS LINE

     %-------------------------------------------------------------------------
     % The introductory (header) part of the paper
     %-------------------------------------------------------------------------

     % The title of the paper. Use \shorttitle to indicate an abbreviated title
     % for use in running heads (you will need to uncomment it).

\title{Wombat, the high intensity diffractometer in operation at the Australian Centre for Neutron Scattering}
%\shorttitle{Short Title}

     % Authors' names and addresses. Use \cauthor for the main (contact) author.
     % Use \author for all other authors. Use \aff for authors' affiliations.
     % Use lower-case letters in square brackets to link authors to their
     % affiliations; if there is only one affiliation address, remove the [a].

\cauthor[a]{Helen E.}{Maynard-Casely}{helen.maynard-casely@ansto.gov.au} {}
\author[a]{Siobhan M.}{Tobin}
\author[b]{Chin-Wei}{Wang}
\author[a]{Vanessa K.}{Peterson}
\author[a]{James R.}{Hester}
\author[a]{Andrew J.}{Studer}

\aff[a]{Australian Nuclear Science and Technology Organisation, Locked Bag 2001, Kirrawee DC, 2232 \country{Australia}}
\aff[b]{National Synchrotron Radiation Research Center, Hsinchu 300092, \country{Taiwan}}

     % Use \shortauthor to indicate an abbreviated author list for use in
     % running heads (you will need to uncomment it).

%\vita{Author's biography}

     % Keywords (required for Journal of Synchrotron Radiation only)
     % Use the \keyword macro for each word or phrase, e.g. 
     % \keyword{X-ray diffraction}\keyword{muscle}

%\keyword{keyword}

     % PDB and NDB reference codes for structures referenced in the article and
     % deposited with the Protein Data Bank and Nucleic Acids Database (Acta
     % Crystallographica Section D). Repeat for each separate structure e.g
     % \PDBref[dethiobiotin synthetase]{1byi} \NDBref[d(G$_4$CGC$_4$)]{ad0002}

%\PDBref[optional name]{refcode}
%\NDBref[optional name]{refcode}

\maketitle                        % DO NOT DELETE THIS LINE

\begin{synopsis}
Description of the Wombat neutron diffraction instrument, operational for the last 17 years at the Australian Centre for Neutron Scattering.  
\end{synopsis}

\begin{abstract}
Wombat is the high intensity neutron diffractometer in operation at the Australian Centre for Neutron Scattering. While primarily used as a high-speed powder diffractometer, the high-performance area detector allows both texture characterisation and single-crystal measurements.  The instrument can be configured over a large range of operational parameters, which are characterised in this contribution to aid experimental planning.  Wombat is particularly optimised for the study of materials \textit{in situ} and \textit{in operando} using the wide range of sample environment available at the centre.   Over  17 years of operation, Wombat has been used to explore a broad range of materials, including: novel hydrogen-storage materials, negative-thermal-expansion materials, cryogenic minerals, piezoelectrics, high performance battery anodes and cathodes, high strength alloys, multiferroics, superconductors and novel magnetic materials. This paper will highlight the capacity of the instrument, recent comprehensive characterisation measurements, and how the instrument has been utlised by our user community to date. 

\end{abstract}

     %-------------------------------------------------------------------------
     % The main body of the paper
     %-------------------------------------------------------------------------
     % Now enter the text of the document in multiple \section's, \subsection's
     % and \subsubsection's as required.

\section{Introduction}

%\begin{figure}   
%
%    \begin{minipage}[ht]{0.3\textwidth}
%    \includegraphics[width=\linewidth]{TOC_wombat_sign.png}
%    \end{minipage}
%
%    \hspace{\fill}
%
%\caption{Potential TOC graphic, Image Credit Jeff Boyd, CC BY-NC-ND %2.0 }
%\end{figure}

The High-Intensity Diffractometer, otherwise known as ‘Wombat’, at the Australian Centre for Neutron Scattering (ACNS, which was known as the Bragg Institute between 2005--2016) has been operating a user program since 2008 as part of the Australian Nuclear Science and Technology Organisation (ANSTO).  Over the last 17 years data from the instrument has contributed to over 400 reports, theses and journal publications.  In the context of new and upgraded neutron diffraction instruments in the Asia/Oceania region \cite{he2023performance, xu2019physical, nambu2024neutron} we explore here the unique flexibility and capability of Wombat, hoping that this will inspire the neutron community to use the instrument in new and innovative ways.

The instrument is accessible to the Australian and international neutron user community through proposal rounds that are run twice a year, with deadlines usually falling on or around 15\textsuperscript{th} March (for instrument time nominally July-December) and 15\textsuperscript{th} September (for instrument time nominally for January - June).  Success rates of proposals are usually about 50 \%, with 200 instrument days available for user operations each year.  

Wombat is inherently  flexible, allowing for a large range of diffraction characteristics.  This is achieved through the three available monochromators, continuous variation in the instrument take-off angle, and a large area detector which couples with a large sample area for mounting of a wide range of types of samples and sample environments.  Below we outline the instrument layout, processes for data collection and reduction and chart the range of instrument configurations possible. The resolution functions of representative configurations have been characterised to aid experimental planning.  Additionally, a number of case studies that showcase the large range of science applications that can be achieved with the instrument will be presented, as well as a overview of the instrument's user community.  

The instrument was originally described \cite{studer2006wombat} during the construction phase. It is timely now to update the community on the progress that the instrument has made, and provide up to date configuration and performance information from the instrument as it presently operates.  

\section{Instrument layout}

\begin{figure}   
\label{Instrument_layout}
\caption{Schematics of the Wombat instrument. Both diagrams show the instrument configured with a take-off angle of 90$^{\circ}$, and a detector start angle of $\approx$ 14$^{\circ}$, which is the most common configuration of these components.  The upper panel shows the `walk in' view, displaying the relative height of the components, while the bottom panel gives a birds' eye view of the instrument layout with the beam path marked in purple. The scale marker is approximate}
    \begin{minipage}[ht]{0.45\textwidth}
    \includegraphics[width=\linewidth]{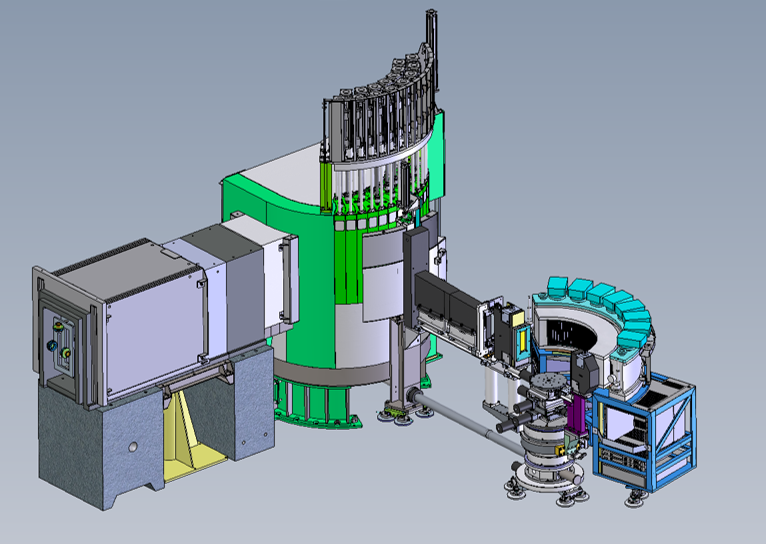}
    \end{minipage}

    \hspace{\fill}

    \begin{minipage}[ht]{0.45\textwidth}
    \includegraphics[width=\linewidth]{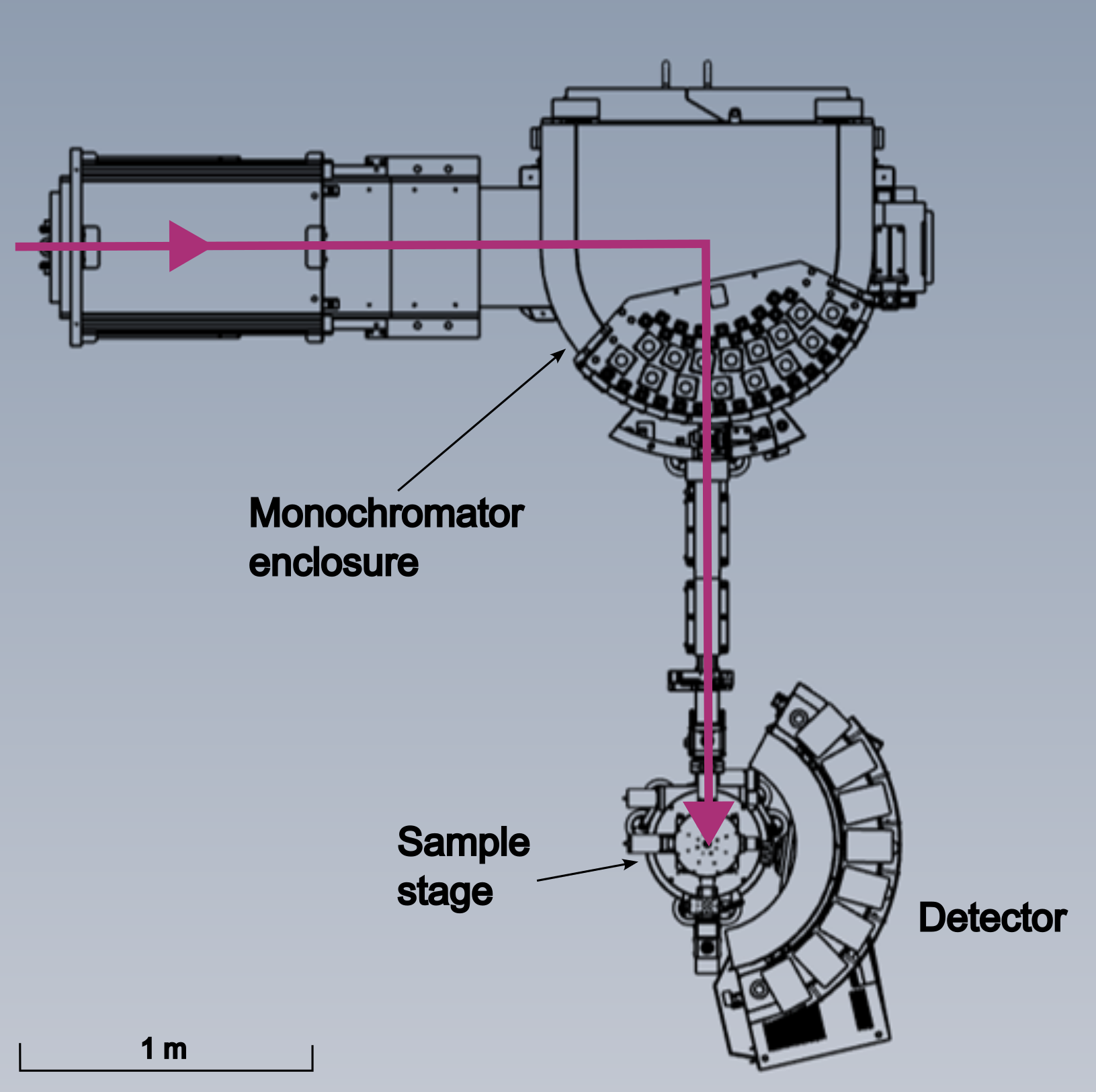}
    \end{minipage}

\end{figure}

The Wombat instrument is located on Thermal Guide 1 from the Open Pool Australian Lightwater (OPAL) reactor, upstream from the High Resolution Powder Diffraction instrument (known as Echidna) \cite{avdeev2018echidna}.  A schematic of the instrument is detailed in Figure \ref{Instrument_layout}, which shows how a thermal neutron guide of internal height 50cm delivers neutrons into a chamber housing the monochromator.  The chamber has been designed with lead shutters that are pneumatically lifted as the take-off angle of the instrument is changed.

The flight path from the monochromator to the sample table has sufficient space to include Helium-3 based neutron polarisers and polarisation analysers \cite{klose2014neutron}, to enable full polarisation analysis for magnetism research on the instrument.  Filters may also be inserted to exclude harmonic contamination incident from the monochromators.  Most commonly used is a highly oriented pyrolitic graphite (HOPG) filter that is regularly paired with the focusing Ge monochromator in the 90$^{\circ}$ take-off position oriented to the 113 reflection (2.41~\AA).  The instrument also has a cooled beryllium filter, which is used in conjunction with the focusing graphite monochromator for wavelengths above 4 \AA. A tapered snout with an adjustable rectangular exit slit is normally placed immediately before the sample position. Typical exit beam dimensions from this snout are 10mm(h) x 20mm(v), with a fully open vertical aperture of around 60mm.

The detector table houses the electronics and high-voltage amplifiers for the instrument, and moves independently around the sample stage to adjust the accessible range of diffraction angles, 2$\theta$. The lowest and highest accessible $2\theta$ angles are $\sim 12^{\circ}$ and $\sim 150^{\circ}$, with a span of 120$^{\circ}$ available at any specific detector position. The sample stage and detector table move together (via a series of air-pads) when the take-off angle of the instrument is changed.  Take-off angle is continuously variable between 60-120$^{\circ}$, but this range is restricted to 75-105$^{\circ}$ when the cooled beryllium filter is in place.  

\section{Data collection and reduction}

\subsection{Detector}
The key component of the Wombat instrument is the detector, a curved position sensitive detector constructed by Brookhaven National Laboratory.  The detector itself is nearly identical to that designed first for the PCS instrument at LANSCE Los Alamos \cite{chen2017fifteen}, which was later moved to the WAND$^2$ instrument at HIFR at Oak Ridge National Laboratory \cite{frontzek2018wand2}.  It consists of eight 15$^{\circ}$ curved panels with a radius of curvature of 728 mm and a height of 200 mm, placed 728 mm from the sample position.  These are constructed such that there is no gap between them, resulting in 120$^{\circ}$ continuous coverage. In standard settings, the detector provides a 968(h)x128(v) pixel image, corresponding to a pixel size of 0.125$^{\circ}$ (h) x 1.6mm(v). The horizontal spatial resolution may be reconfigured to 0.01625$^{\circ}$ if required. The two-dimensional detector means that samples throughout the continuum of crystallite orientational order can be studied, ranging from well-randomised powders, through to textured samples and single crystals.  On Wombat, this detector is used in conjunction with a radial collimator constructed by JJ X-ray, which has a 0.5$^{\circ}$ pitch, excluding parasitic scatter from a radius \textgreater50 mm from the instrument centre.  

Instrument control and data collection is undertaken with the SINQ Instrument Control Software \cite{heer1997sinq}, which is usually accessed through the GumTree interface \cite{lam2006gumtree}.

\subsection{Data reduction}\label{Data:reduction}

For many years, data reduction at the instrument was carried out using the Large Array Manipulation Program (LAMP) \cite{richard1996analysis}, and macro routines are still available for use with this program.  Since 2020 there has been a transition to data reduction implemented using Python code within the Gumtree framework \cite{lam2006gumtree}.  
The Python code is freely available \cite{hesterwombatreduction}.

A typical raw dataset from Wombat contains one or more frames of data, representing counts accumulated across the detector in some
time period. The data reduction steps described below are those most commonly used when deriving one-dimensional datasets suitable
for input into typical powder diffraction analysis software; other types of measurements typically require alternative approaches to data reduction, for example using Int3D \cite{katcho2021int3d} for single-crystal measurements.

\subsubsection{Normalisation}

Beam monitor counts are used to place each 
frame of data onto a uniform scale, as the incident intensity from the reactor
source is likely to fluctuate. Multiple datasets may be treated simultaneously to place them on the same scale.

\subsubsection{Efficiency correction}

Each pixel on the detector has a variable response to an incident neutron. To account for
this, an efficiency correction map is created by measuring a flood field from a Vanadium rod,
then subtracting a background measured without the rod. In order to account for the presence of
weak coherent peaks from Vanadium, the detector angle $stth$ is stepped by 0.125$^{\circ}$ over a range
of about 12$^{\circ}$ during both rod and background
measurements, which typically takes an entire day. After subtraction, any pixels judged by a configurable automatic algorithm to contain peak contributions are removed from the final
calculation. After averaging the response for each retained pixel, the final efficiency map is
used to calculate a correction factor for each pixel, which is applied at this step. To account for gradual
changes in the response pattern of the detector, new efficiency maps are generated every few months.

\subsubsection{Vertical integration}\label{vertint}

The final step in data reduction is summation of all pixels at each nominal $2\theta$ position to produce
a 1D diffraction pattern suitable for input into powder diffraction modeling software. Prior to this summation, users
may optionally remap all pixels to their `true' $2\theta$ position, which is termed 'straightening' in the present paper.
This procedure invalidates the asymmetry model of Finger, Cox and Jephcoat \cite{finger1994correction} and may therefore complicate diffraction pattern peak-shape modeling, as the calculation of true $2\theta$ assumes a parallel beam and point scatterer, which is rarely the case in most experiments. Nevertheless, in practice the resulting peaks are often sufficiently well-modelled using standard pseudo-Voight profile parameters. 

\subsubsection{Gain re-refinement}\label{gain}
Where the sample contains  strong incoherent scatterers, the efficiency correction described above is often insufficient for the high background. In such cases, several frames of data can be collected by stepping the detector
by the pixel spacing (typically 0.125$^\circ$), resulting in multiple measurements of the same signal at each angle after vertical summation. Scale factors for each column
of vertical pixels are then refined using the procedure of Ford and Rollett \cite{fordrollett}. Wombat data reduction uses the same underlying code as Echidna \cite{avdeev2018echidna}, which should be referred to for details of the mathematics involved.

The final 1D diffraction pattern is one output of this refinement. If straightening has been requested (see \S\,\ref{vertint}), the final 1D diffraction pattern is instead obtained
by using the refined scale factors to rescale the 2D frames before straightening and vertical summation.

\subsubsection{File output}

Data may be output in two variants of three-column ASCII format suitable for GSAS-II \cite{toby2013gsas} and Topas \cite{coelho2018topas}, as well as a pdCIF file \cite{tobypdcif} containing comprehensive metadata, including locally-defined
data names capturing all data reduction parameter settings. It is generally possible to exactly reproduce data reduction based on the contents of the pdCIF file.

\section{Instrument configurations and characterisations}

To assist with understanding published Wombat studies and experimental planning, each of the three monochromators used on
Wombat has been characterised with appropriate standard materials and resolution curves determined for a variety of take-off angles.    

\subsection{Resolution function fitting}
Data were measured from standard materials contained in 6 mm diameter vanadium cans and reduced as described above with no straightening applied. Raw
datasets and reduced data are available on Zenodo \cite{studer_pgmono, maynard_casely_ge335, wang_ge115}.

Data were then analysed within the TOPAS5 refinement suite \cite{coelho2018topas}.  Resolution functions were fitted with the following procedure: first the wavelength, sample displacement and zero offset error for a given monochromator reflection and take-off angle were determined from a refinement of LaB$_6$ standard reference material 660b, supplied by the National Institute of Standards and Technology, USA \cite{black2011certification} which has been isotopically enriched with an 11B precursor \cite{Lab6_660b}.  For these refinements, a Chebyshev 5-term background and scale factor were also refined.  No atomic positions were refined, but the isotropic displacements of both lanthanum and boron atoms were.   

\begin{figure} % wombat resolution curves
\centering
    %\begin{minipage}[ht]{0.45\textwidth}
    \includegraphics[width=13cm]{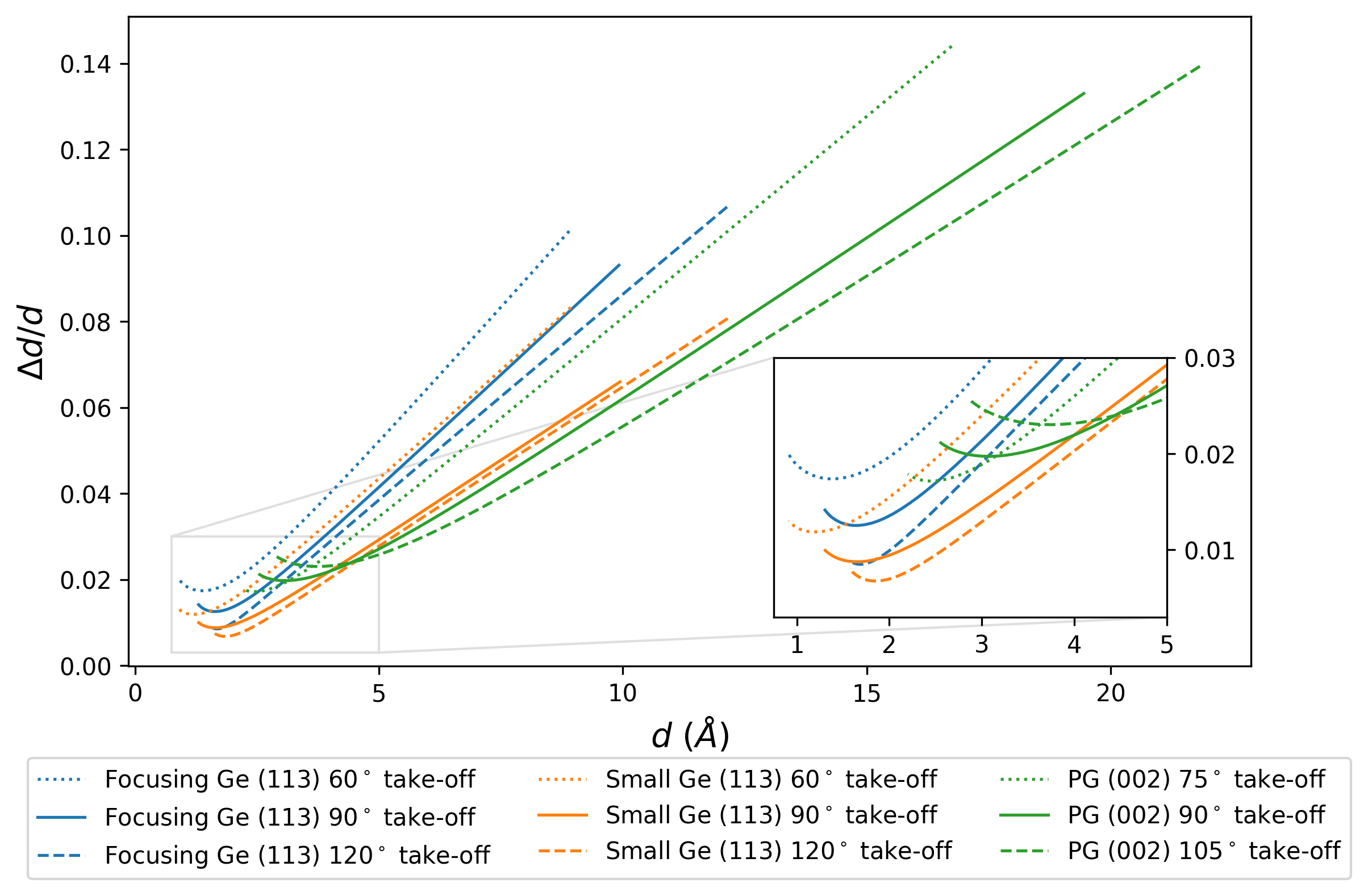}
    %\end{minipage}
    %\hspace{\fill}
\caption{$\Delta d/d$ resolution curves calculated from the peak shape functions fitted for the 113 reflection of the focusing Ge monochromator, 113 reflection of the flat Ge monochromator and 002 reflection for the focusing graphite monochromator across the range of take-off angles.}
\label{fig:res_curves}
\end{figure}

Once established, the wavelength and zero offset error values were used to undertake Pawley refinements of patterns of Na$_2$Ca$_3$Al$_2$F$_{14}$ (NAC, sourced from Laboratoire des Oxydes et Fluorures, Facult\'e des Sciences Universitie du Maine, France) collected at each configuration. Although NAC is not an official standard reference material, it has been used as a line profile standard \cite{courbion1988na2ca3al2f14}.  Given Wombat's broad range of operaterating wavelengths (1-6 \AA) it was chosen as a consistent standard material to characterise peak shape and asymmetry across all configurations. The sample of NAC is also mixed with CaF$_2$, which was also accounted for in the refinements.  For all of the configurations, the peak shape was fit with a Gaussian peak-width variation using a simple three-parameter $UVW$ model derived from the Caglioti equation \cite{caglioti1958choice}:

\begin{equation}
FWHM = (U\,\tan^2\theta + V\,\tan \theta + W)^{1/2}
\end{equation}

where $\theta$ is the Bragg angle, that is, half of the diffraction angle $2\theta$.  No crystal size parameter was added, but lattice parameter was allowed to refine. An overview of the $\Delta d/d$ resolution curves calculated from the fitted $U$, $V$ and $W$ parameters for the 113 reflection of the focusing Ge monochromator, 113 reflection of the flat Ge monochromator and 002 reflection for the focusing graphite monochromator across the range of take-off angles is plotted in Figure \ref{fig:res_curves}.

The large aperture of the position-sensitive detector (over 200 mm of height) means that asymmetry is prominent in low angle peaks.  The approach to fitting the asymmetry is detailed in the next sections. 

\subsubsection{Focusing Ge monochromator}

The `workhorse' monochromator of the instrument is a 300 mm vertically focusing monochromator, which consists of 23 Germanium crystal plates that each have a length of 70 mm cut along the [115] direction, with the [1$\bar{1}$0] direction vertical.  As [1$\bar{1}$0] is vertical and the Ge space group is cubic, the $\langle hhn\rangle$ directions lie in the horizontal plane, allowing $h$$h$$n$ reflections to be rotated into reflecting position.  The focus of the monochromator can be changed, but the optimum between intensity and divergence is $\sim$ 0.3$^{\circ}$ curvature, hence this was fixed throughout the collection of the data from the standards.   Peak shapes were fit for the two main reflections used, 113 and 115, over the range of take-off angles detailed in Table \ref{tab:focus Ge UVW}. The majority of the configurations described in Table \ref{tab:focus Ge UVW} could be fit without any additional asymmetry corrections. The exception was at the higher take-off angle for the 115 reflection where the asymmetry becomes more pronounced due to peaks appearing at lower $2\theta$ angles.  For this configuration the Finger \textit{et al.} model \cite{finger1994correction} was used with the `filament' length ($S$) parameter set at 20 mm to reflect the incident neutron slit size and the receiving slit length ($D$) parameter set at 200 mm to reflect the size of the detector.

%%%%%%%%%%%%%%%%%%%%%%%%%%% focus Ge UVW fits %%%%%%%%%%%%%%%%%%%%%%%%%%

\begin{table}
\caption{Fitted $U$, $V$ and $W$ parameters, along with wavelengths (determined from LaB$_6$ refinements) refined from data of a NAC sample for the two main reflections used of the focusing Ge monochromator, across a range of take-off angles.  As described in the text, fitting of the 115 reflection at 120$^{\circ}$ required additional modeling for the peak asymmetry ($S$ and $D$). Flux relative to Ge 115 at 90$^{\circ}$ is indicative only; * indicates pyrolitic graphite filter inserted}
\begin{tabular}{lllllllll}      % Alignment for each cell: l=left, c=center, r=right
Configuration   & $\lambda$ (\AA) & Relative flux &  U & V & W  & S & D & wRp (\%)  \\
\hline
113 reflection 90$^{\circ}$  & 2.417 & 2.5$^*$ & 0.965(14) & -0.98(4) & 0.534(18)  & - & - & 3.953  \\
115 reflection 90$^{\circ}$  & 1.543 & 1.0 & 0.88(3) & -1.15(5) & 0.63(2)  & - & -  & 3.354 \\
113 reflection 60$^{\circ}$  & 1.706 & 2.1 & 1.63(6) & -1.00(8) & 0.39(2)  & - & -  & 3.564  \\
115 reflection 60$^{\circ}$  & 1.089 & 0.3 & 1.8(5) & -1.5(3) & 0.53(5)  & - & -  & 3.120  \\
113 reflection 120$^{\circ}$  & 2.958 & 1.8 & 0.43(2) & -0.70(5) & 0.64(2)  & - & -  & 4.834  \\
115 reflection 120$^{\circ}$  & 1.890 & 1.0 & 0.50(2) & -1.09(5) & 0.92(2)  & 20 & 200 & 3.022 \\
\end{tabular}
\label{tab:focus Ge UVW}
\end{table}

%%%%%%%%%%%%%%%%%%%%%%%%%%%%%%%%%%%%%%%%%%%%%

\subsubsection{Flat Ge Monochromator}

In addition to the two monochromators originally described \cite{studer2006wombat}, Wombat has routinely used a single cut flat (335) Germanium monochromator, a heritage component from the MRPD instrument previously operated on the HiFAR reactor \cite{howard1994neutron}.  The use of this monochromator has proven  advantageous for experiments requiring higher flux than that available on Echidna, but with more resolution than is possible with the focusing Ge monochromator.  As such it is now used in $\sim$10 \% of Wombat experiments.  Three reflections are routinely used from this monochromator: 113, 115, and 224. The peak shape parameters of these reflections across a range of take-off angles are given in Table \ref{tab:flat Ge UVW}.  The asymmetric contribution to peak shape is more pronounced due to the narrower instrumental resolution, so the Finger \textit{et. al.} model was used, with the sample length set at 20 mm and receiving slit length set at 200 mm as described in the previous section.

%%%%%%%%%%%%%%%%%%%%%%%%%%% flat Ge UVW fits %%%%%%%%%%%%%%%%%%%%%%%%%%

\begin{table}
\caption{Fitted $U$, $V$ and $W$ parameters, along with wavelengths (determined from LaB$_6$ refinements) refined from data of a NAC sample for the three reflections used regularly from the flat Ge monochromator, across a range of take-off angles.  In each fit the Finger \textit{et al.} asymmetry model was applied with $S$ = 20 mm and $D$ = 200 mm. Flux relative to Ge 115 at 90$^{\circ}$ is indicative only.}
\begin{tabular}{lllllll}      % Alignment for each cell: l=left, c=center, r=right
Configuration   & $\lambda$ (\AA) & Relative flux &  U & V & W  & wRp (\%) \\
\hline
113 reflection 90$^{\circ}$  & 2.422 & 0.3 & 0.48(4) & -0.50(7) & 0.267(3)  & 4.938  \\
115 reflection 90$^{\circ}$  & 1.544 & 0.2 & 0.38(1) & -0.38(1) & 0.235(7)  & 3.803  \\
224 reflection 90$^{\circ}$  & 1.638 & 0.3 & 0.904(2) & -1.07(1) & 0.47(2)  & 3.313  \\
113 reflection 60$^{\circ}$  & 1.713 & 0.3 & 0.70(4) & -0.49(7) & 0.25(1)  & 4.683   \\
115 reflection 60$^{\circ}$  & 1.093 & 0.1 & 0.6(6) & -0.5(5) & 0.2(4)  & 4.646  \\
224 reflection 60$^{\circ}$  & 1.159 & 0.2 & 1.17(5) & -0.96(7) & 0.36(1)   & 3.337 \\
113 reflection 120$^{\circ}$  & 2.962 & 0.3 & 0.368(19) & -0.58(4) & 0.38(2)  & 6.592   \\
115 reflection 120$^{\circ}$  & 1.891 & 0.2 & 0.290(17) & -0.53(3) & 0.37(1)   & 5.117  \\
224 reflection 120$^{\circ}$ & 2.005 & 0.3 & 0.374(7) & -0.700(7) & 0.474(4)   & 5.538   \\
\end{tabular}
\label{tab:flat Ge UVW}
\end{table}

%%%%%%%%%%%%%%%%%%%%%%%%%%%%%%%%%%%%%%%%%%%%%

\subsubsection{Focusing graphite monochromator}

A highly-oriented pyrolytic graphite monochromator, constructed identically to the focusing Germanium monochromator and cut to the 002 reflection, may also be used on Wombat.  One use originally envisioned for this monochromator was a `very' high flux 2.4 \AA \,\,option at a take-off angle of 42$^{\circ}$.  In practice this has been rarely used, as the flux with the focusing Ge monochromator 113 reflection at 90$^{\circ}$ has been sufficient for all rapid experiments to date.  However, the graphite monochromator has proven  useful in providing a longer wavelength, particularly for magnetic studies, and is used in $\sim$\,5\,\% of Wombat experiments.  To characterise these configurations, which would have  few incident peaks from LaB$_6$, NAC was used to refine lattice parameters as well as subsequent line profiles. These measurements were taken in tandem with a cooled beryllium filter, which limits the accessible take-off angles. The Cagliotti peak shape parameters fit for the 002 reflection over the range of accessible take-off angles are detailed in Table \ref{tab:graphite UVW}.  As for the focusing Ge monochromator, it was found that a satisfactory peak shape could be fit from all the configurations without the need for an additional asymmetry model.  

%%%%%%%%%%%%%%%%%%%%%%%%%%% Big Graphite UVW fits %%%%%%%%%%%%%%%%%%%%%%%%%%

\begin{table}
\caption{Fitted $U$, $V$ and $W$ parameters, along with wavelengths refined from NAC sample data for the 002 reflection of the focusing graphite monochromator, across a range of take-off angles. Flux relative to Ge 115 at 90$^{\circ}$ is indicative only.}
\begin{tabular}{lllllll}      % Alignment for each cell: l=left, c=center, r=right
Configuration  & $\lambda$ (\AA)  & Relative flux & U & V & W  & wRp (\%)   \\
\hline
002 reflection take-off 75$^{\circ}$  & 4.109(3) & 13  & 2.92(2) & -2.39(9) & 1.21(2) & 2.085   \\
002 reflection take-off 90$^{\circ}$  & 4.779(1) & 11 & 2.0(1) & -1.7(2) & 1.06(8)  &  1.813  \\
002 reflection take-off 105$^{\circ}$  & 5.349(2) & 9.0 & 1.6(1) & -1.9(2) & 1.4(1)  & 2.219   \\
\end{tabular}
\label{tab:graphite UVW}
\end{table}

%%%%%%%%%%%%%%%%%%%%%%%%%%%%%%%%%%%%%%%%%%%%%

\subsection{Sample environment configurations}\label{sample_env}
\begin{figure} % wombat sample environment popularity plot
\centering
    %\begin{minipage}[ht]{0.45\textwidth}
    \includegraphics[width=\textwidth]{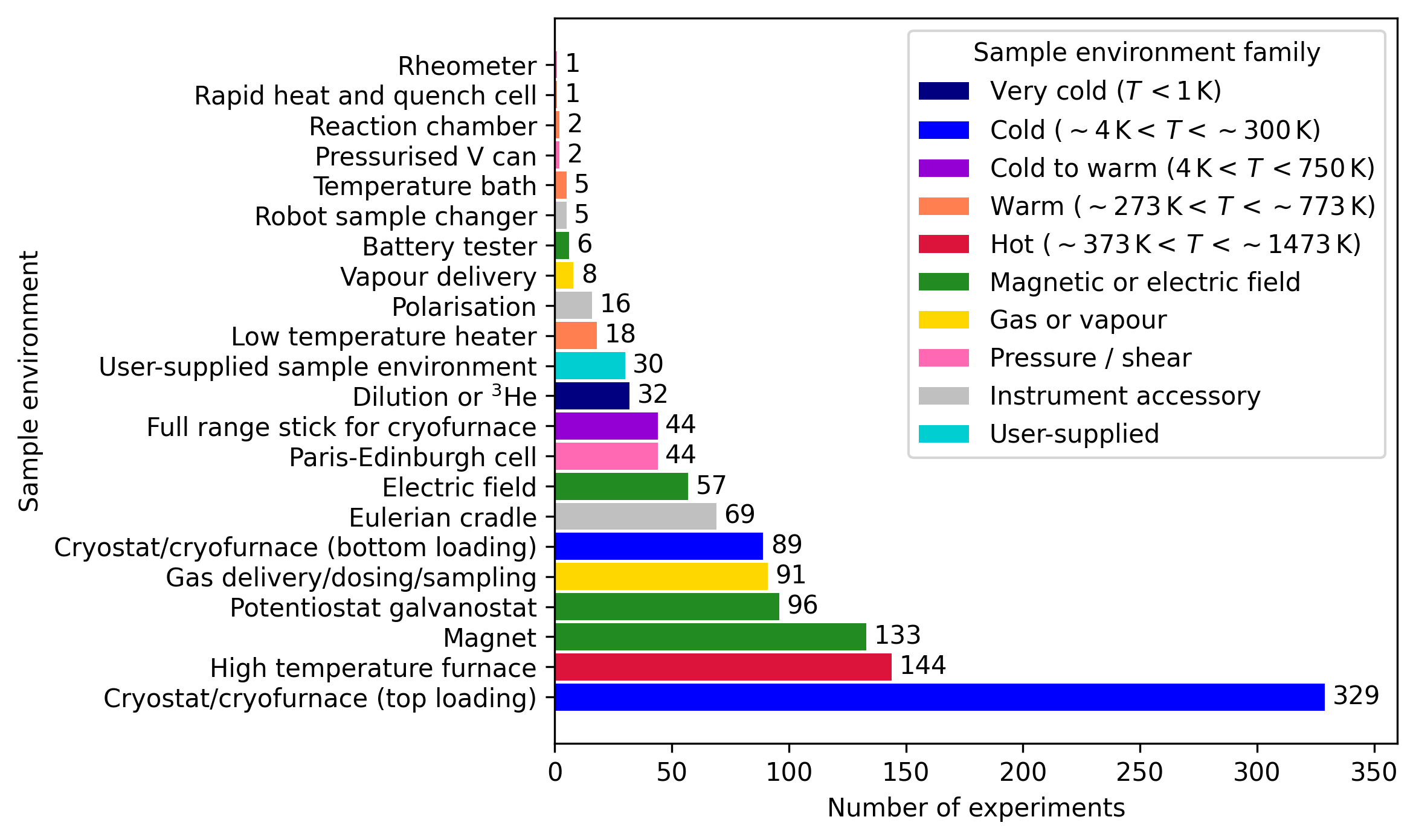}
    %\end{minipage}
    %\hspace{\fill}
\caption{Bar chart illustrating the use of different pieces of sample environment equipment on Wombat during user experiments.  Multiple pieces of sample environment equipment may be used during a single experiment, which is why the total sample environment equipment uses charted (1222) exceeds the number of experiments (805) listed in Section \ref{User_community}. }
\label{fig:sample_environment_bar_chart}
\end{figure}

Wombat's generously-sized sample stage can host a huge variety of sample environments, spanning the breadth and depth of many phase diagrams. Some families of sample environment equipment include:
\begin{itemize}
    \item Cold temperature 4~K$\, \lesssim T \lesssim\, 300$~K (cryostats, cryofurnaces), 
    \begin{itemize}
        \item Including also  cold $T < 1\,$K (dilution insert and a one-shot $^3$He insert that was previously available),
    \end{itemize}
    \item Warm temperature $273\,$K $\lesssim\,T\,\lesssim\,773\,$K (temperature baths, low temperature heaters and furnaces),
    \begin{itemize}
    \item To span the temperature range from cold to warm $4\,$K$\, < T < 750\,$K, a `full range' sample stick can be used in the cryofurnaces,
    \end{itemize}
    \item Hot temperature 373~K $\lesssim T \lesssim\,1473$~K (high temperature vacuum furnace),
    \item Magnetic field, electrical field and charge variation (cryomagnets $\leq$\,11\,T, electric field generation $\leq$\,10\,kV, battery tester, potentiostat galvanostat),
    \item Gas handling (gas dosing, gas condensing, gas sampling, vapour delivery),
    \item and Pressure / shear (pressurised sample cans, Paris-Edinburgh press $\leq$\,10\,GPa, rheometer).
\end{itemize} 
The popularity of various sample environment configurations can be gauged from the bar chart shown in Figure~\ref{fig:sample_environment_bar_chart}.  Users are also able to supply their own sample environment, subject to safety and feasibility requirements; for example, an induction furnace has been developed by a group from Newcastle University, details of which can be found in \citeasnoun{merz2023complex}.

Additional instrument accessories such as the Eulerian cradle, polarisation capability and a robotic sample changer further extend the flexibility of the instrument.  The Eulerian cradle pairs particularly well with Wombat's large area detector, enabling studies of texture and single crystal samples (further discussed in Section \ref{Use:non-powder}). While most sample environments listed in Figure~\ref{fig:sample_environment_bar_chart} are best suited to powder samples, some of the cryofurnaces and the cryomagnet have a sample probe rotation $\omega$ axis, which facilitates aligned single crystal work as will be described in Section \ref{Use:non-powder}. 

%%%%%%%%%%%%%%%%%%%%%%%%%%%%%%%%%%%
% Wombat experiments section
%%%%%%%%%%%%%%%%%%%%%%%%%%%%%%%%%%%
\section{Wombat `on the run': examples of experiments on the instrument} \label{applications}

Like other high-flux neutron diffractometers, Wombat has been used predominately for \textit{in situ} studies, and as presented in Section \ref{sample_env} the majority of these have been in variable temperature conditions.  The highly configurable nature of the instrument means that multiparameter experiments can also be undertaken. These have included pairing of magnetic field variation with temperature (for example \citeasnoun{heinze2019magnetic} ), gas dosing with temperature (for example \citeasnoun{auckett2018continuous}) and a novel high-pressure high-electric field experiment \cite{yamane2019search}.  High flux additionally affords the study of small samples, such as those recovered from high-pressure high-temperature synthesis; Wombat regularly measures multi-phase samples as small as 50 mg, such as those presented in \citeasnoun{brown2024synthesis}.   

Selected novel experimental setups and approaches on the instrument have been developed to the point that they are a regular offering as part of the Wombat user program. These are described in the following sections.    

\subsection{Rapid data collection}

One application envisaged when designing Wombat was rapid real-time acquisition, a result of combining the high-flux source with the high-count rate capacity of the detector.  Wombat has come to specialise in two types of rapid data collection experiments: stroboscopic measurements where the triggering of the collection is synchronised to external sources, and `one-shot' experiments where rapid data is acquired on triggering to monitor real-time changes to a sample.  

An example of a stroboscopic measurement is the synchronisation of the detector collections with a pulse generator to perform high frequency cyclical measurements, initially of piezoelectric materials \cite{pramanick2010situ}.  With this setup 250\,Hz measurements with multiple time bins are routine.  This approach has been extended to the study of multiferroic materials at realistic operating frequencies \cite{hinterstein2023stroboscopic}.

\begin{figure}   

    \begin{minipage}[ht]{0.8\textwidth}
    \includegraphics[width=\linewidth]{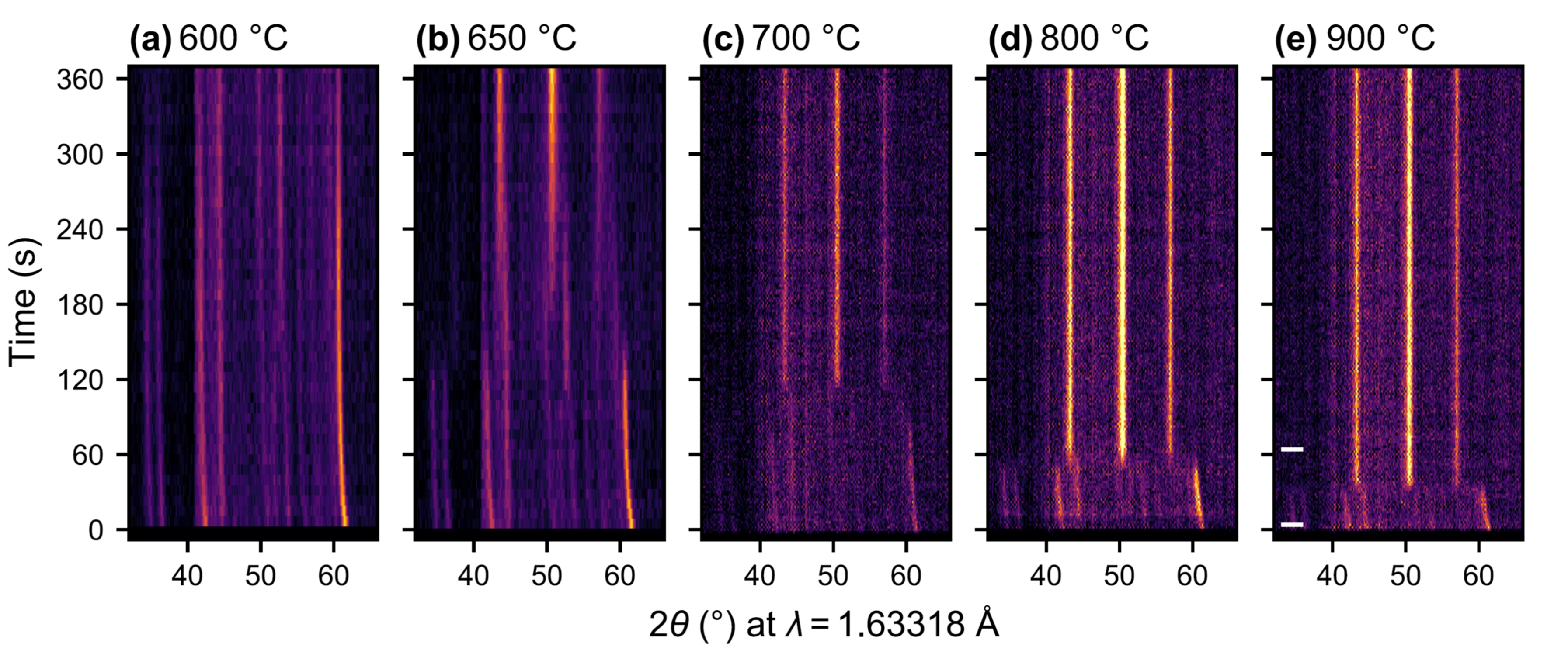}
    \end{minipage}

    \hspace{\fill}

\caption{Neutron diffraction patterns collected in situ as reagents mixed for Na$_{0.7}$WO$_3$ are dropped into a preheated furnace, set to various temperatures. The colourmap limits are identical across (c–e) but are different from those in (a, b) due to the different acquisition conditions. These patterns are shown with the SiO$_2$ background subtracted.  Reprinted with permission from \citeasnoun{tegg2021intermediate}. Copyright 2021 American Chemical Society.}
\label{fig:speedy}
\end{figure}

One-shot experiments have been undertaken on Wombat to elucidate the response of material in response to rapid heating.  Collecting with continuous acquisition means that the data can be segregated down to 50\,ms, with no deadtime between frames.  This has been used to investigate the response of materials on timescales ranging from 2\,s per pattern \cite{tegg2021intermediate} (Figure \ref{fig:speedy}) to 0.05\,s per pattern \cite{merz2023complex}.

\subsection{Hydrogenous powder diffraction collections}

\begin{figure}   

    \begin{minipage}[ht]{0.9\textwidth}
    \includegraphics[width=\linewidth]{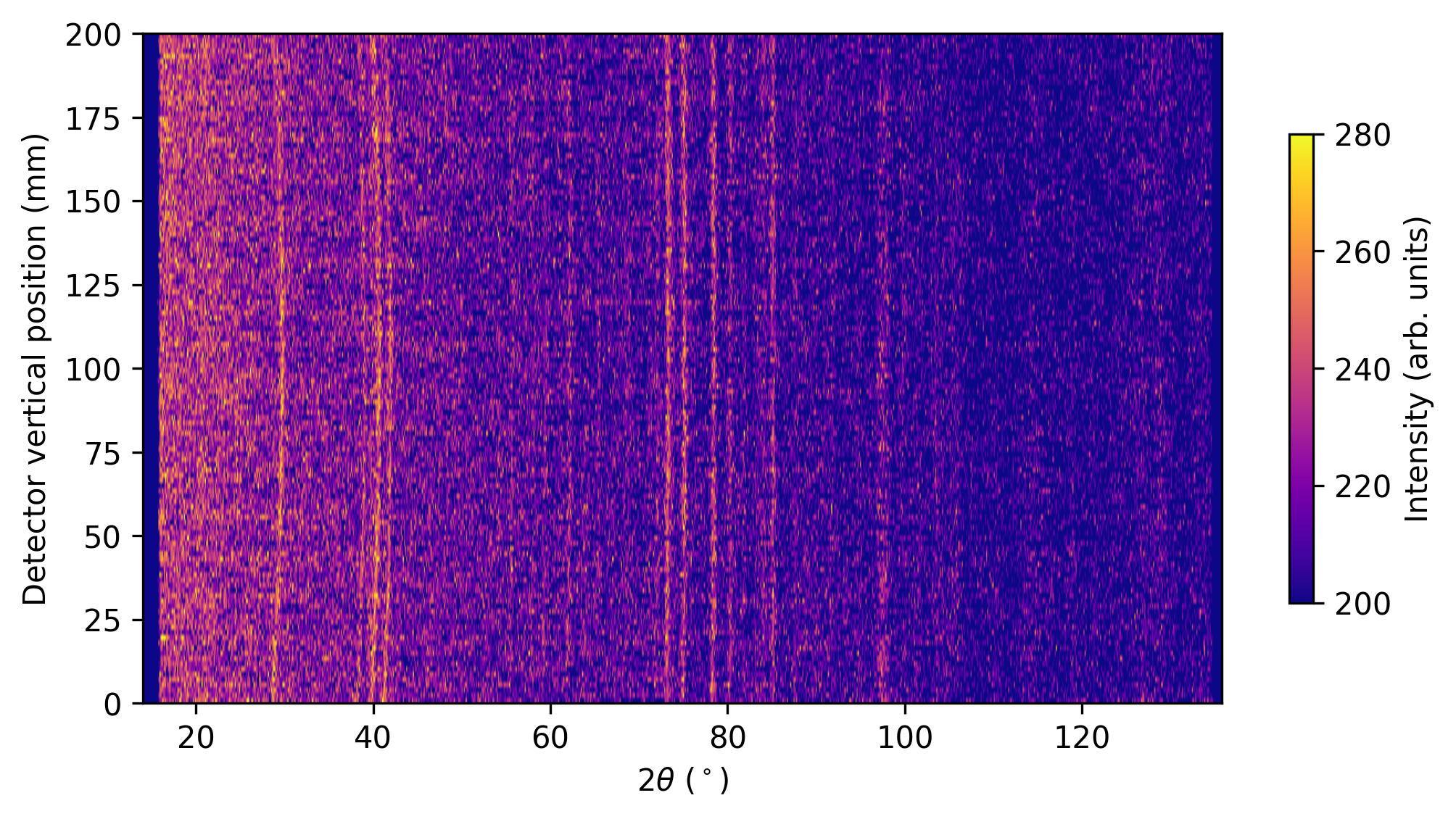}
    \end{minipage}

    \hspace{\fill}

\caption{Detector image from a recent data collection of C$_3$H$_8$ propane on Wombat, collected at $\lambda$ = 2.422 \AA~and processed using the gain refinement strategy explained in the text.  This illustrates that despite the low symmetry ($P21/n$ space group) and high hydrogen content the Bragg peaks can be seen over the high incoherent background. }
\label{fig:propane}
\end{figure}

The common approach when studying hydrogenous materials with neutron powder diffraction is to substitute the hydrogen atoms with deuterium, lowering the background arising from the large incoherent scattering length of hydrogen atoms.  This approach has been successful, but there will be some cases where deuteration is not feasible.  The high flux of the Wombat instrument affords an opportunity to work around this, and to undertake diffraction studies directly from hydrogenous materials.  For example, a study of solid methane on the Wombat instrument \cite{maynard2020re} allowed lattice parameters to be measured between 4-82\,K and the thermal expansion fitted. Data collection from these high-hydrogen content materials is improved by scanning the detector over a range of angles and running gain re-refinement as detailed in \S\,\ref{gain}, with an example presented in Figure \ref{fig:propane}.  This has been in turn applied to a range of small molecular systems with a number of applications \cite{klapproth2019kinetics, choi2020phase, maynard2022crystal, chen2023lowering}.  

\subsection{\textit{In operando} collections}

In contrast to \textit{in situ} measurements, some experiments wish to capture the behaviour of materials within real-world technology. An example of this is the response of battery materials, where an accurate understanding of component function cannot be fully gained using diffraction data of electrode powders or batteries held at particular states of charge. Hence \textit{in operando} experimental methods have been developed, time-resolved measurements of materials undergoing change while within a functioning device \cite{peterson2024neutron}. 

\begin{figure}   

    \begin{minipage}[ht]{\textwidth}
    \includegraphics[width=\linewidth]{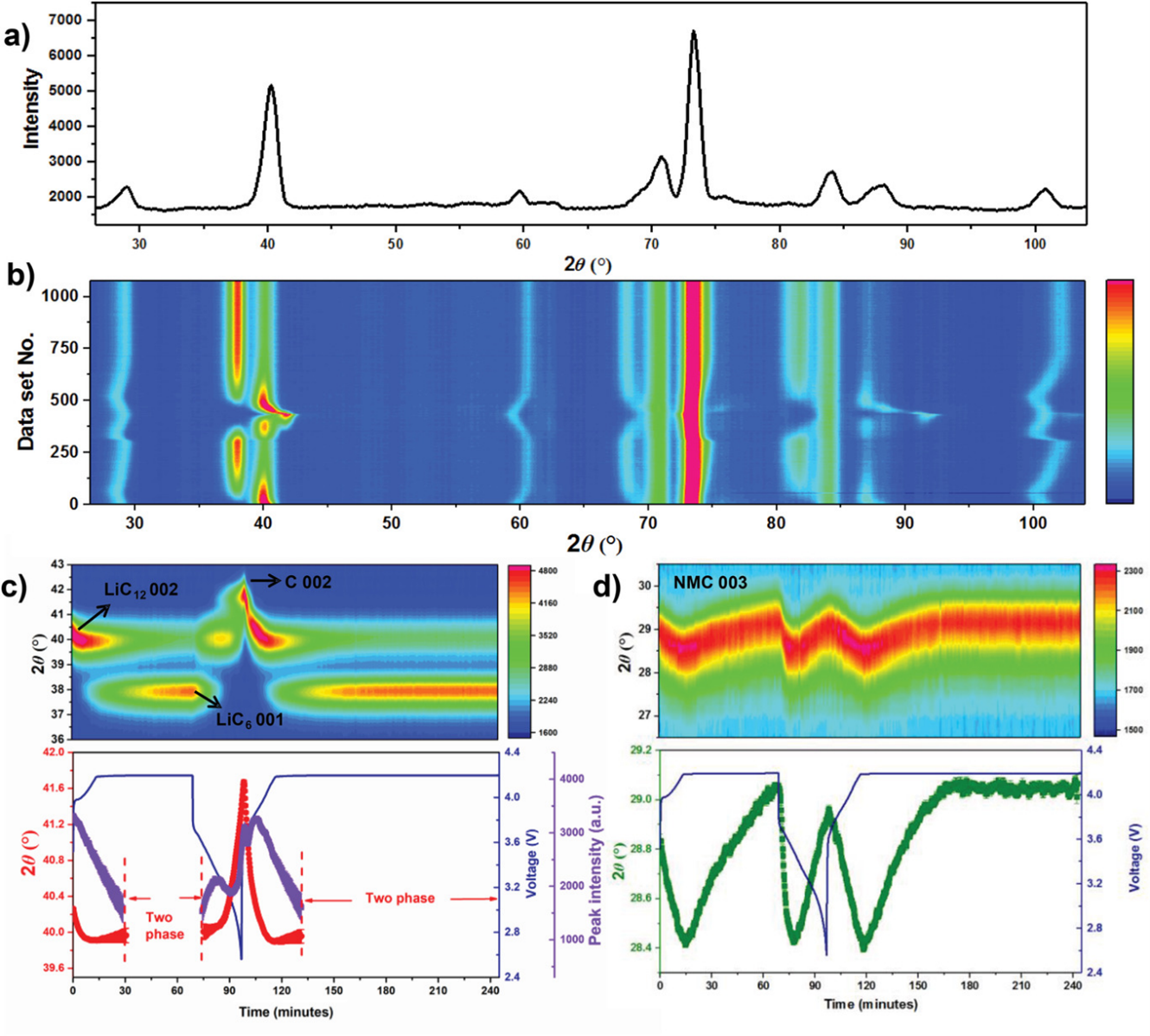}
    \end{minipage}

    \hspace{\fill}

\caption{a) The first 10 s dataset for an \emph{in operando} series of diffraction data of a battery b–d) Top: contour plot of data with intensity in color (legend shown right). c) and d) results of Gaussian peak-fitting of a negative (c) and positive (d) electrode reflection. Reproduced with permission from \citeasnoun{liang2020battery}, all data collected at $\lambda$ = 2.4168(1) \AA.}
\label{fig:in_operando_battery}
\end{figure}

The first \textit{in operando} diffraction experiment of a lithium ion battery was conducted on Wombat, with data acquired every 5 min during battery cycling \cite{sharma2010battery}. This work showed that the phase evolution of the battery electrodes could be studied while the battery was being used. All diffraction data represent a time-averaged measurement, necessitating a consideration of the distribution of states for the material undergoing structural change in the analysis and interpretation of any particular dataset. Following the 2010 study, \textit{in operando} diffraction studies of batteries have increased in number, and evolved to enable a more detailed understanding of electrode function, necessary to inform improvements to performance, as well as economic and environmental sustainability \cite{liang2020battery}. An important factor for \textit{in operando} battery measurements is the time in which useful (Rietveld quality) data can be gained; in operando data of a commercial battery have been collected on Wombat with a 10 s time resolution \cite{liang2020battery}, Figure \ref{fig:in_operando_battery}.

The \textit{in operando} approach had also been applied to using Wombat to study adsorbent materials used in the separation and storage of gases \cite{peterson2017sorbent}. The \textit{in operando} conditions allow study of gas incorporation and interaction within an adsorbent material in an environment that is representative of the real-world application.

\subsection{Non-powder data collections} \label{Use:non-powder}

Though Wombat was originally designated as a high-intensity powder diffractometer, its large-area position-sensitive detector has proven advantageous for volumetric surveying of reciprocal space for both single-crystal and texture diffraction experiments.  The volume of accessible reciprocal space can be extended when an Eulerian cradle is mounted on the sample stage, allowing sample axis angles $\omega$, $\chi$, and $\phi$ equivalent to the Eulerian angles to be measured. The cradle can be used both to undertake measurements and to manipulate crystal orientation for subsequent study within other sample environments. Comparing our Eulerian cradle four-circle geometry to that of Busing \& Levy, using the same Cartesian co-ordinate system defined in their paper \cite{businglevy1967}: our geometry has all angles ($2\theta, \, \omega, \chi,\, \phi$) as right-handed rotations about their respective axes, whereas the system of Busing \& Levy has $2\theta, \, \omega, \phi$ as left-handed, and $\chi$ as right-handed.

\begin{figure}   

    \begin{minipage}[ht]{0.9\textwidth}
    \includegraphics[width=\linewidth]{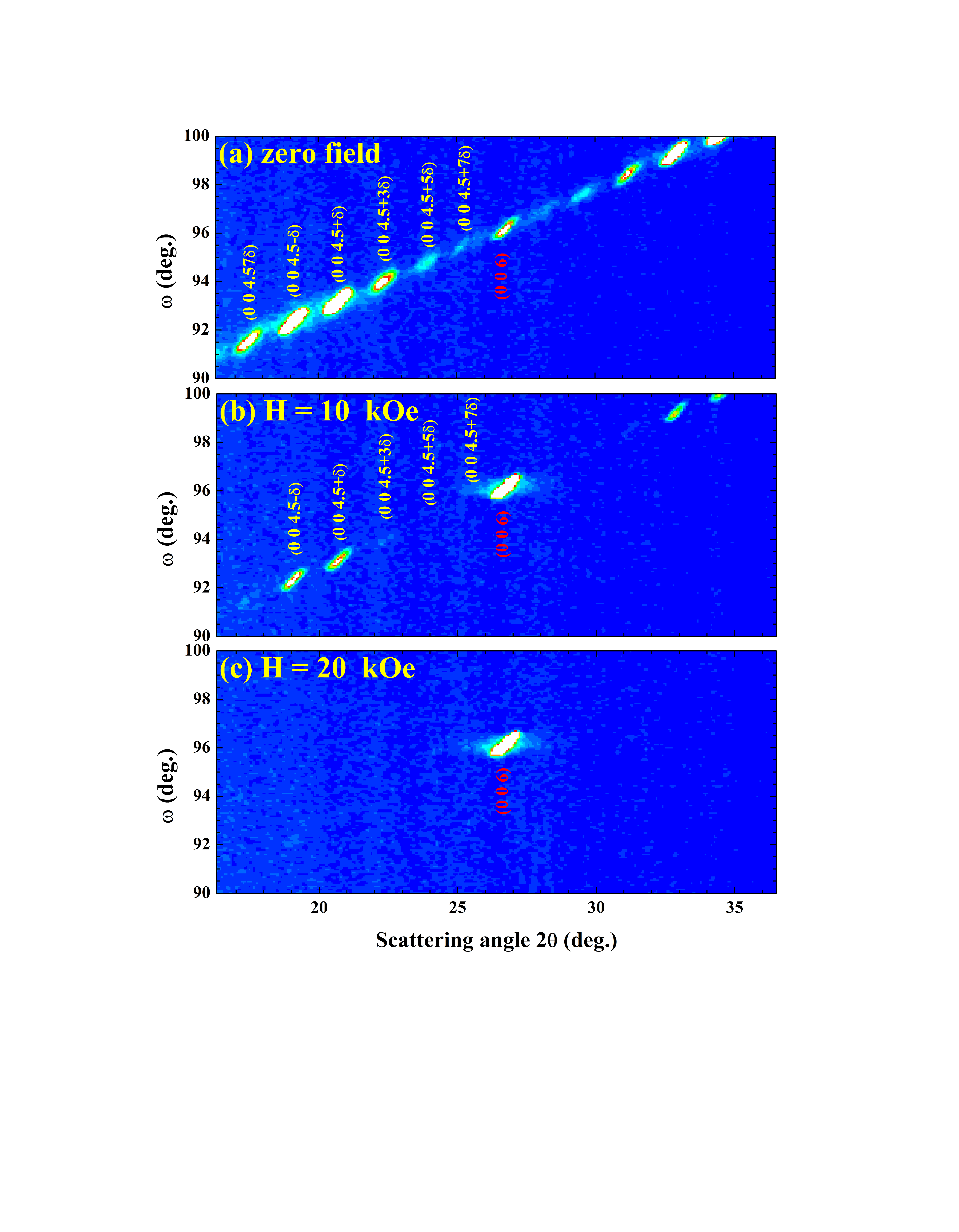}
    \end{minipage}
    
    \hspace{\fill}
\caption{Plotted from data presented in \citeasnoun{gao2024high}, charting the response of a single crystal of the triangular-lattice magnet HoPdAl$_4$Ge$_2$ to different magnetic field strengths, collected at $\lambda$ = 2.41 \AA. Interested readers are referred to that paper for the interpretation of these results.}
\label{fig:single_crystal}

\end{figure}

At ACNS, measurement of large volumes of reciprocal space for single crystals is
usually carried out using the Laue diffractometer Koala \cite{piltz2018koala}, due to the relative
speed of the Laue technique. For the majority of single-crystal measurements on Wombat, 
only a small region of reciprocal space is studied, with the sample aligned and mounted so that the normals to the lattice planes of interest coincide with the scattering plane of the instrument. The data acquisition is performed through a step-scan in the azimuthal angle ($\omega$), for which the rotation axis is perpendicular to the horizontal scattering plane. The typical increment of $\omega$ used in Wombat measurements is 0.1$^{\circ}$ or 0.2$^{\circ}$. As Wombat's detector can receive out-of-plane scattering with a span of $\pm7.8^{\circ}$, a sample with a large out-of-plane lattice spacing allows a survey of reciprocal space that simultaneously covers multiple parallel reciprocal lattice planes.
Changes in this volume of reciprocal space can be efficiently monitored as a function of sample
environment variable, such as temperature or pressure. This experimental setup is
regularly used to obtain data from single crystals as a function of magnetic field strength,
with field direction perpendicular to the scattering plane, as illustrated in Figure \ref{fig:single_crystal}. 

Such a setup is advantageous in studying spin-flip, spin-flop, or other metastable phase transitions in magnetic materials \cite{gao2024high}.   Recently, the Int3D software package \cite{katcho2021int3d} has been deployed on the instrument to enhance the analysis of single-crystal diffraction data collected on Wombat, allowing diffraction spots to be indexed and their integrated intensities extracted.  

The vertical extent of the detector also allows for the rapid mapping of orientation space for texture analysis, in contrast to point-detector techniques that scan single peaks in succession.  This has been undertaken with both single peak fitting and whole profile methodologies \cite{simons2014measurement, xu2021multiple}.  Orientational effects are significant in multiferroic materials, and Wombat has undertaken \textit{in situ} experiments to understand distributions in the static \cite{lu2016electric} and dynamic \cite{hinterstein2023stroboscopic} structural response to electric fields.  

%%%%%%%%%%%%%%%%%%%%%%%%%%%%%%%%%%%
% Wombat user community section
%%%%%%%%%%%%%%%%%%%%%%%%%%%%%%%%%%%
\section{The Wombat user community}\label{User_community}

Our Wombat instrument is part of a vast global ecosystem of central facility instruments, which serve a diverse user community with far-reaching impacts, as outlined in the neutron research landscape surveys of \citeasnoun{velichko2025rendering} and \citeasnoun{shukla_major_2022}.  As such, it is good to understand how an individual instrument is contributing to this ecosystem, and that support is being maintained over the timescale of user operations.  In this section, all analysis and discussion is based on data from the merit-based access proposal rounds and director's discretionary beamtime for the period 2008--2023. In 2024 there was a long shutdown ($\sim$~6 months) of the OPAL reactor, so we discount user statistics from 2024 as the availability of Wombat was much lower than usual.  From 2008 -- 2023 Wombat was utilised in 805 experiments across 670 proposals. This includes single experiment and program proposals awarded through merit-based access proposal rounds, as well as director's discretionary beamtime. In addition to this, a number of commercial experiments have taken place, where a commercial client paid for access to the instrument. Furthermore, Wombat instrument scientists and other ANSTO staff are not included in the user demographic data.

\begin{figure} % wombat user community time series
\centering
    %\begin{minipage}[ht]{0.45\textwidth}
    \includegraphics[width=10cm]{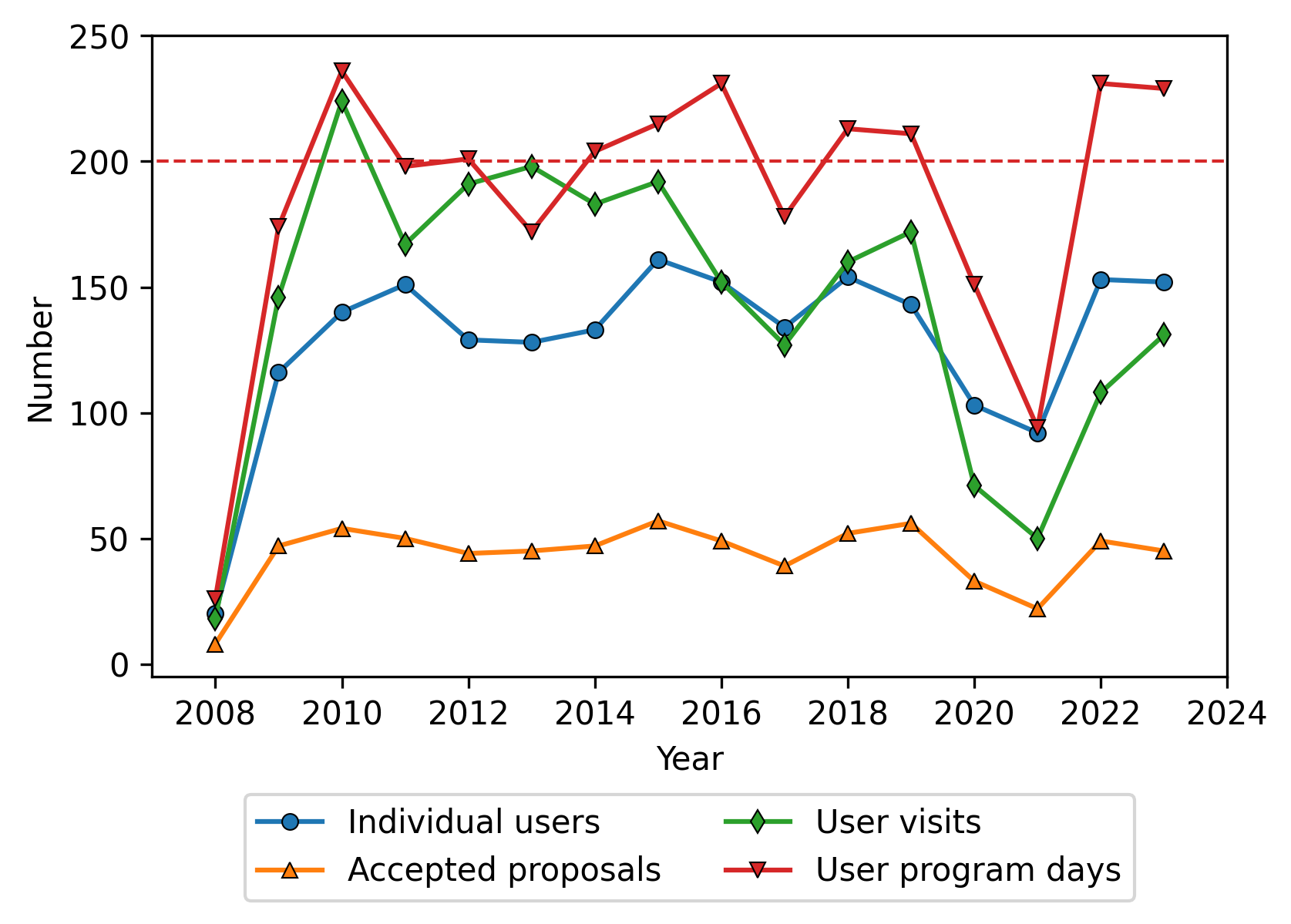}
    %\end{minipage}
    %\hspace{\fill}
\caption{Annual Wombat instrument use 2008--2023, including number of individual users, number of user visits, number of accepted proposals and number of user program days. The target of 200 days for the user program is indicated by the red dashed line.}
\label{fig:users_vs_year}
\end{figure}

The Wombat user community has grown to encompass 1031 individuals who have collectively made 2290 visits to the instrument. Figure~\ref{fig:users_vs_year} shows how much Wombat was used each year. 2008 was the year the instrument was commissioned. Thereafter the use of Wombat increased dramatically before reaching a period of steady state operation 2010--2019: $\sim$~200 user program days, $\sim$~50 accepted proposals, and $\sim$~140 individual users per annum. ACNS aims to provide 200 days of user access to each instrument annually. This is underpinned by a  reliable neutron source, the multi-purpose OPAL research reactor which is at power $\sim$~300 days per year. 

The COVID-19 pandemic during 2020--2021 disrupted usual programming for Wombat, as it did for everyone else, with restrictions on working on-site and travel for users. Since then, normal operations have resumed, although total user visits are slightly down. A fully remote user program is not offered by the ACNS; however Wombat's sibling, Echidna, offers a mail-in service for short proposals run in limited sample environment conditions.

%%%%%%%%%%%%%%%%%%%%% proposal duration table
\begin{table}
\caption{Table of duration of experiments associated with different proposal types.}
\begin{center}
\renewcommand{\arraystretch}{1.5}
\begin{tabular}{m{3cm} m{1.7cm} m{2cm} m{3cm} m{3cm} }      % Alignment for each cell: l=left, c=center, r=right
\hline
 Proposal type    & Number of proposals        & Number of experiments     & Mean duration of experiments (days) & Standard deviation of duration of experiments (days) \\
\hline
Normal & 561 & 679 & 3.6 & 1.8 \\
Director's discretionary & 64 & 67 & 3.6 & 1.6 \\
Program & 45 & 59 & 4.9 & 2.2 \\
\hline
\end{tabular}
\end{center}
\label{tab:prop_duration}
\end{table}
%%%%%%%%%%%%%%%%%%% end table

\begin{figure} % wombat days per proposal histogram
\centering
    %\begin{minipage}[ht]{0.45\textwidth}
    \includegraphics[width=10cm]{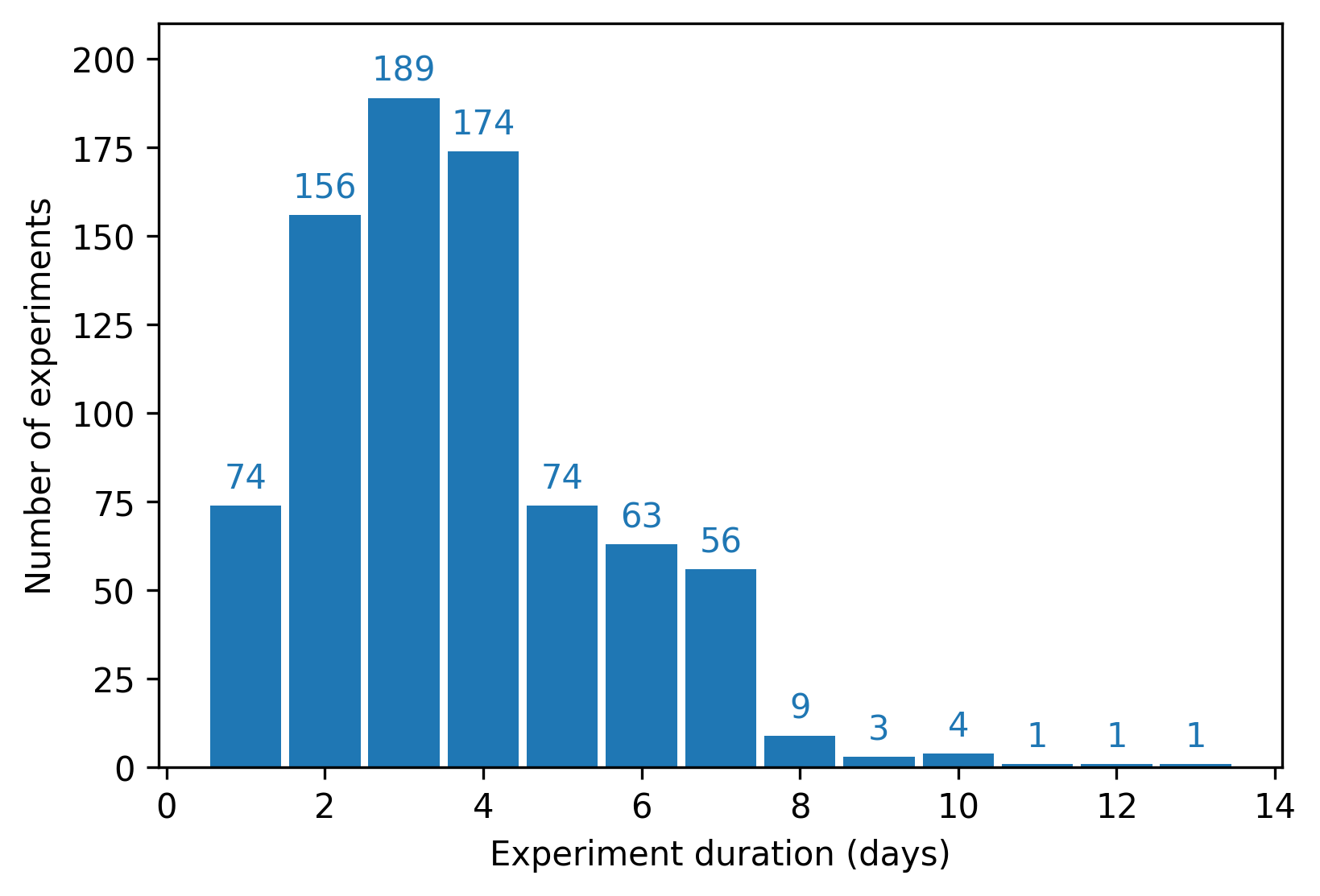}
    %\end{minipage}
    %\hspace{\fill}
\caption{Duration of experiments on Wombat. This histogram presents data associated with normal, director's discretionary beamtime, and program proposals.}
\label{fig:proposal_days_histogram}
\end{figure}

The general trend indicated by Figure~\ref{fig:users_vs_year} is for there to be two or three visiting users and around four days allocated to each proposal. Further analysis of accepted proposals (Figure~\ref{fig:proposal_days_histogram} and Table~\ref{tab:prop_duration}) shows that four days is indeed typical for normal proposals and director's discretionary beamtime, which usually involve one single experiment on the instrument. The amount of time awarded to director's discretionary beamtime is consistent with the aim that no more than 10\,\% of user program days should be allocated via this route. Program proposals tend to entail multiple experiments on Wombat and other instruments, and usually involve more days of beamtime overall (typically five days per experiment). 

\begin{figure} % wombat users positions/organisations bar chart
\centering
    %\begin{minipage}[ht]{0.45\textwidth}
    \includegraphics[width=12cm]{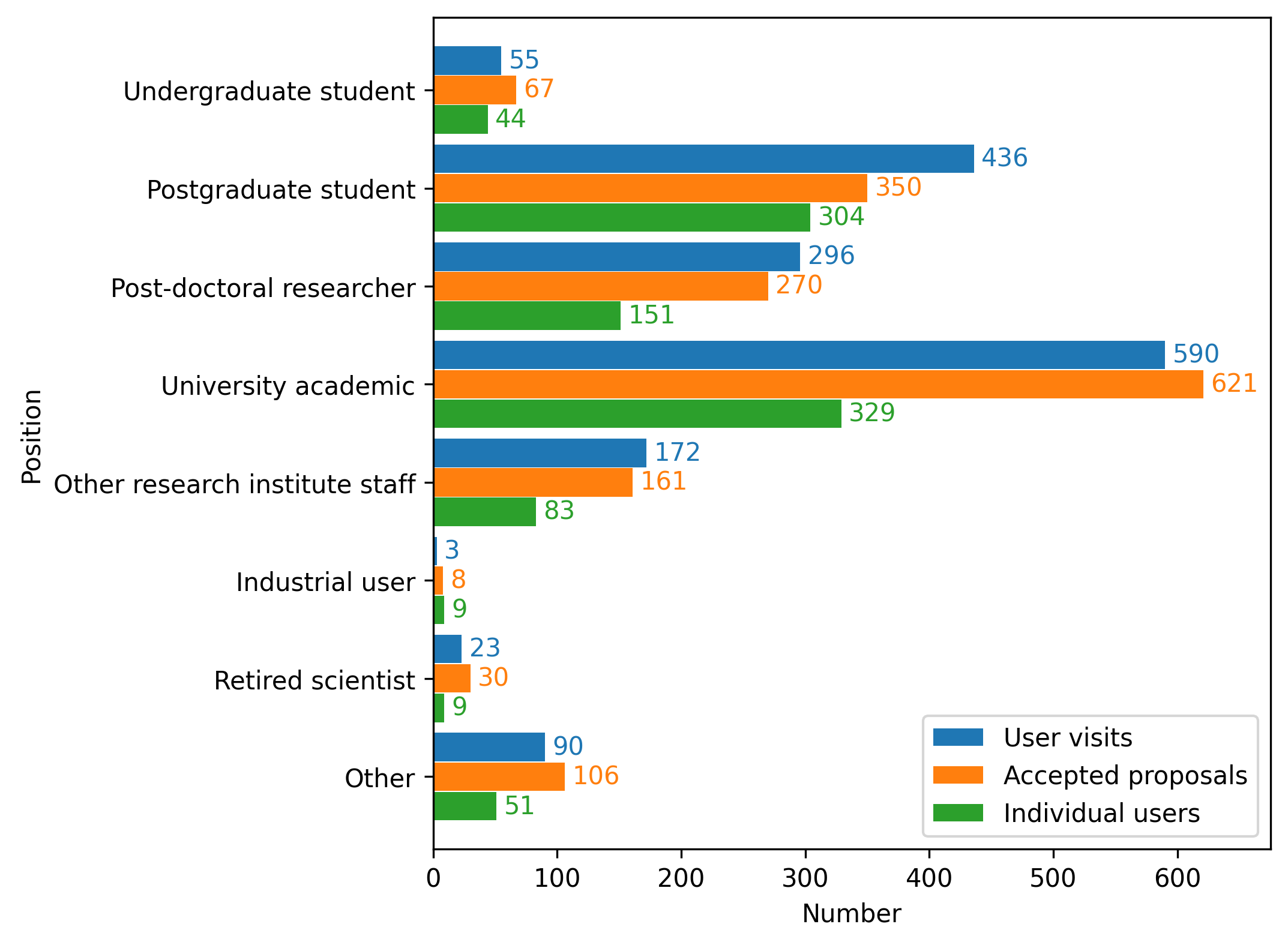}
    %\end{minipage}
    %\hspace{\fill}
\caption{The Wombat user community according to position/organisation. Here an accepted proposal is counted if it features \emph{at least one} user from that category, hence the number of proposals on this chart exceeds the total number of unique proposals 2008--2023. As people progress through their careers, their position changes and so some individual users appear in more than one category in this chart. Industrial users are those who have accessed the instrument through the merit-based proposal rounds. Commercial clients who have paid for access to the instrument are not included. ANSTO staff are not included.}
\label{fig:user_position_bar_chart}
\end{figure}

The contribution of Wombat to the wider neutron scattering community is made plain to see in Figure~\ref{fig:user_position_bar_chart}, which displays the career stages or positions of Wombat users. The vast majority of Wombat users hail from universities, with the instrument supporting the work of 329 academics, 151 post-docs and 304 PhD students from 2008--2023. PhD students seem more likely to attend experiments in person, relative to accepted proposals, than other groups. The distribution of users at all career stages, from undergraduate student right through to retired scientist, highlights Wombat's contribution to a thriving research ecosystem based around central facilities such as neutron and synchrotron beamlines.

\begin{figure} % wombat publications each year
\centering
    %\begin{minipage}[ht]{0.45\textwidth}
    \includegraphics[width=10cm]{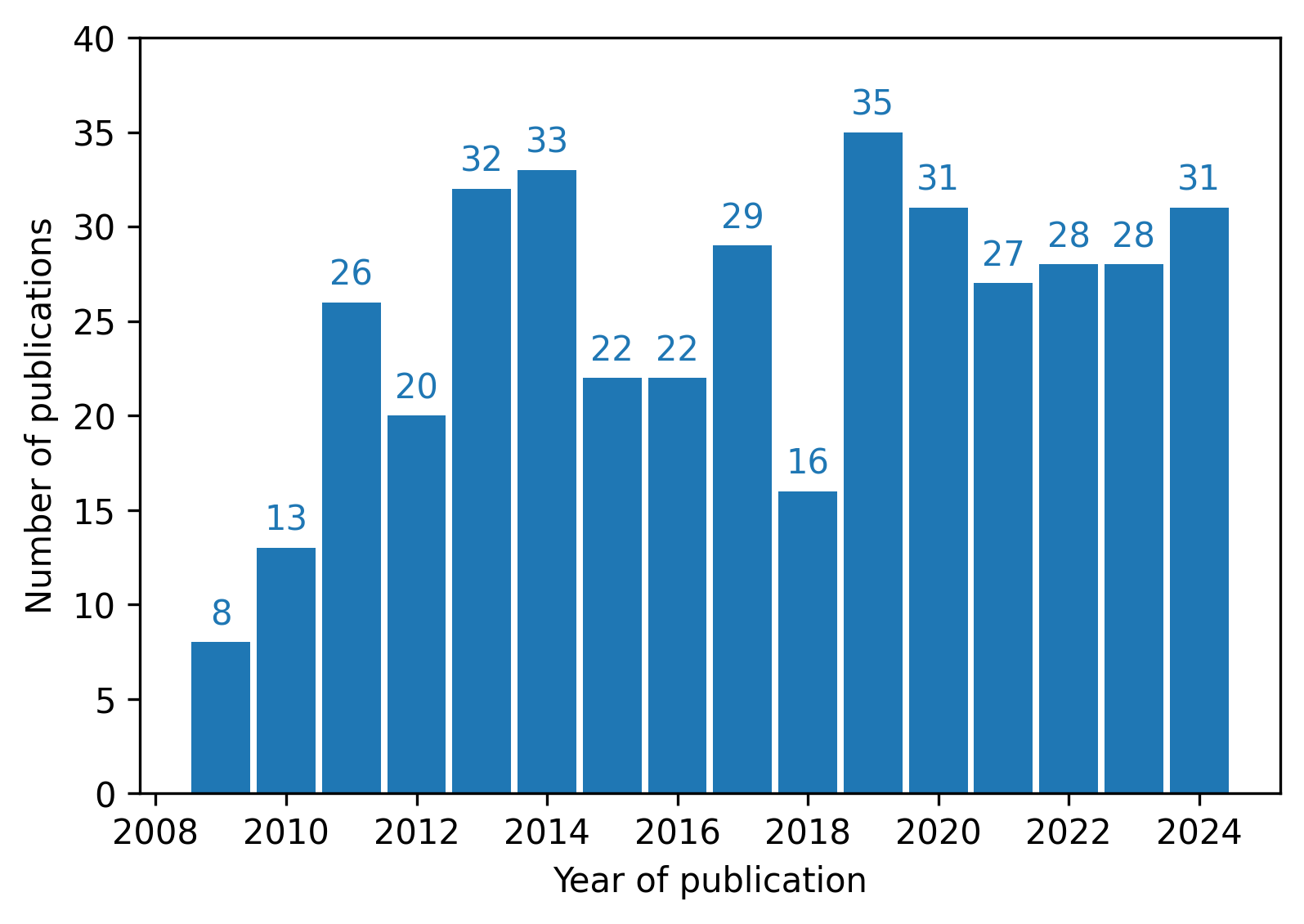}
    %\end{minipage}
    %\hspace{\fill}
\caption{Annual publications featuring results from Wombat.}
\label{fig:publications_per_year}
\end{figure}

Ultimately, central facility instruments are usually judged on their peer-reviewed output. Wombat experiments have resulted in 402 peer-reviewed publications from 2008 to 2024 inclusive. Figure~\ref{fig:publications_per_year} charts this across the years of operation, showing a consistency in output of between 20 -- 30 papers a year. This demonstrates that the instrument has achieved a sustainable output, which reflects the accessibility and quality of the data for preparing publications against the inherent experimental challenges presented by most of our user experiments.      

%%%%%%%%%%%%%%%%%%%%%%%%%%%%%%%%%%%
% Summary section
%%%%%%%%%%%%%%%%%%%%%%%%%%%%%%%%%%%
\section{Summary and the future of Wombat}

Our Wombat instrument is named after an Australian marsupial, and metaphorically refers to the broader `fatter' diffraction peaks, and the speed of the instrument, as wombats can run at a `high intensity' of nearly 40 km/h \cite{wombat_fast} (over short distances).  The Wombat instrument is remarkably configurable and able to undertake unique and sometimes pioneering experiments.  The instrument supports a healthy and diverse user community and we are much indebted to our users who return and challenge the instrument setups year on year.  The Wombat team are always seeking to support new experiments and user teams, and encourage those who are intrigued by the capability of the instrument to get in contact. 

As detailed in Section \ref{applications}, the instrument team support a wide range of scientific applications, and stands ready to use the instrument in support of new scientific directions.  This undertaking will be supported in the long term by the construction of a new detector for the instrument.  With the aim to re-produce the success of the original BNL detector, the new detector will be built in-house at ACNS, an exciting development for detector science in Australia.  We hope to be commissioning this new detector in 2028, and that with this the Wombat instrument will continue to provide excellent neutron scattering results for many years.

     %-------------------------------------------------------------------------
     % The back matter of the paper - acknowledgements and references
     %-------------------------------------------------------------------------

     % Acknowledgements come after the appendices

\ack{Acknowledgments}

The Wombat team would like to recognise the technicians and engineering teams who not only constructed the instrument, but have kept it in good order and providing lots of neutrons to our users over the years: the sample environment team, the electrical engineering team, the data acquisition electronics team, the mechanical engineering team, the health physics team, and the control and software engineers at ACNS.  The propane data presented in Figure \ref{fig:propane} was a test measurement as part of proposal 18728 where Dr Anna Engle was the principal investigator. The Wombat team would like to acknowledge Jason Christoforidis for support in producing the Wombat schematics, Dr Max Avdeev for obtaining the NAC sample, Dr Jamie Schulz for provision of the Wombat usage statistics and Dr Andrew Whitten for the information on Wombat publications.

%% Note added by Overleaf: If using bibtex, remove the "references" environment above, and uncomment the following line.
\bibliographystyle{iucr}
\bibliography{refs}

\end{document}